\documentclass[12pt]{article}
\usepackage{psfig}
\begin{document}

\def\lesssim{\mathrel{\mathpalette\vereq<}}
\def\gtrsim{\mathrel{\mathpalette\vereq>}}
\makeatletter
\def\vereq#1#2{\lower3pt\vbox{\baselineskip1.5pt \lineskip1.5pt
\ialign{$\m@th#1\hfill##\hfil$\crcr#2\crcr\sim\crcr}}}
\makeatother

\newcommand{\rem}[1]{{\bf #1}}

\renewcommand{\thefootnote}{\fnsymbol{footnote}}
\setcounter{footnote}{0}
\begin{titlepage}
\begin{center}

\hfill    UCB-PTH-98/31\\
\hfill    LBNL-41890\\
\hfill    hep-ph/9806218\\
\hfill    \today\\

\vskip .5in

{\Large \bf 
Study of Inclusive Multi-Ring Events\\
from Atmospheric Neutrinos\footnote
{This work was supported in part by the U.S. 
Department of Energy under Contracts DE-AC03-76SF00098, in part by the 
National Science Foundation under grant PHY-95-14797.  HM was also 
supported by Alfred P. Sloan Foundation.}
}

\vskip .50in

Lawrence J. Hall and Hitoshi Murayama

\vskip 0.05in

{\em Department of Physics\\
     University of California, Berkeley, California 94720}

\vskip 0.05in

and

\vskip 0.05in

{\em Theoretical Physics Group\\
     Ernest Orlando Lawrence Berkeley National Laboratory\\
     University of California, Berkeley, California 94720}

\vskip .5in

\end{center}

\vskip .5in

\begin{abstract}

The current analysis of atmospheric neutrinos by the Super-Kamiokande 
Collaboration is based only on fully-contained one-ring events and 
partially contained events.  We show that the up-down ratio of 
fully-contained, inclusive,
multi-ring events gives an independent test of the 
atmospheric neutrino anomaly, without the need for particle identification.  
Moreover, this class of events is rich 
in neutral current events and hence gives crucial information for 
discriminating between oscillations of $\nu_{\mu}$ into 
$\nu_{e, \tau}$ and $\nu_{s}$.

\end{abstract}
\end{titlepage}

\renewcommand{\thepage}{\arabic{page}}
\setcounter{page}{1}
\renewcommand{\thefootnote}{\arabic{footnote}}
\setcounter{footnote}{0}

\section{Introduction}

Recent results from the Super-Kamiokande collaboration have confirmed
that the measured neutrino fluxes, produced by cosmic ray showers in the 
Earth's atmosphere, do not agree with expectations from theoretical
calculations. An anomaly is seen in events with both low visible energy,
the sub-GeV data \cite{SKsub}, and high visible energy, the multi-GeV
data \cite{SKmulti}. 
The collaboration has used three event topologies to present evidence for
these anomalies: fully contained (FC) 1 ring events, where the
Cerenkov ring has the characteristics of an electron or muon --
$e$-like and $\mu$-like events; and partially contained (PC) events,
which Monte Carlo calculations show have a 98\% probability to be
produced by a $\nu_\mu$ charged current interaction. The $e$-like and
$\mu$-like events are studied both in the sub-GeV and multi-GeV
samples, while the PC events are multi-GeV events. From the observed
number of these events, the $\mu/e$ ratio, relative to Monte Carlo 
expectations, is measured to be $0.61 \pm 0.03 \pm
0.05$ for sub-GeV data, and $0.66 \pm 0.06 \pm 0.08$ for multi-GeV data. The 
uncertainties are statistical and systematic, respectively. In the
sub-GeV case the $\mu/e$ ratio is the ratio of 1-ring $\mu$-like to 
1-ring $e$-like events, while in the multi-GeV case the numerator also 
includes the PC events.

The Super-Kamiokande collaboration has also measured the zenith angle
distribution of these event classes, for both sub- and multi-GeV
data. The $e$-like events were consistent with Monte Carlo
expectations, while the distributions for $\mu$-like and PC events
were far from Monte Carlo expectations. Let $\rho_i(\phi)$ be the number of
events of class $i$ with visible products moving in the upward
direction relative to the number of these events in the downward
direction. The visible products must be moving in a cone about the
vertical with half angle $\phi$. These up-down ratios are important,
as they are insensitive to much of the theoretical uncertainties
associated with calculations of the neutrino fluxes from cosmic ray
showers; however, they are only useful if there is a good angular correlation
between the neutrino and the visible products of the neutrino
interaction. For the multi-GeV data, the correlation between neutrino
and charged lepton directions is good, with a mean of 15--20$^o$. 
For the sum of $\mu$-like and PC events, 
Super-Kamiokande has measured $\rho_\mu(78^o) = 0.52 \pm 0.07 \pm 0.01$, 
(Monte Carlo expectation: $0.98 \pm 0.03 \pm 0.02$), while from the
$e$-like events they measure $\rho_e(78^o) = 0.84 \pm 0.13 \pm 0.02$
(Monte Carlo expectation: $1.01 \pm 0.06 \pm 0.03$),
with uncertainties strongly dominated by statistics. While the measured 
value of $\rho_e$ is consistent with Monte Carlo calculations, 
the discrepancy in $\rho_\mu$ is 
especially significant. For the sub-GeV data, the correlation between
neutrino and charged lepton directions is much poorer -- with a mean
of about 60$^o$ -- and the data give $\rho_\mu(53^o) = 0.69 \pm 0.08$ and 
$\rho_e(53^o) = 1.23 \pm 0.12 $, where only statistical uncertainties are 
shown.

This data, together with that of previous experiments,
provides substantial evidence that neutrinos produced in the
atmosphere are oscillating as they traverse the Earth. The low
systematic uncertainty on $\rho_\mu$, both from theoretical calculations
of the flux and from the Super-Kamiokande detector, make it clear that
the survival probability for a $\nu_\mu$ to traverse the Earth is
substantially less than unity. We are therefore led to study other
signals in the Super-Kamiokande data which could
\begin{itemize}
\item {\it Confirm oscillations with low systematic uncertainties.}

Even though large uncertainties cancel from the $\mu /e$ ratio,
several remain: the flux calculation, the charged current cross
section, the neutral current cross section, the energy scale and the
separation of 1-ring from multi-ring events each have systematic
uncertainties of 4--6\% \cite{SKmulti}. On the other hand, it is striking
that the systematic uncertainties for the up-down ratios
$\rho_{e, \mu}$ are much smaller: the flux calculations, detector
asymmetries and downward going muon flux each give systematic
uncertainties of only 1--2\% \cite{SKmulti}. 
The anomaly in the $\nu_\mu / \nu_e$ flux ratio depends crucially on the 
particle identification, which has been quite
convincingly tested at KEK \cite{KEKtest}, but still one would like to
have a test independent of the systematic issues of particle ID.

\item {\it Probe the flavor of the oscillation mode.}

 The measurement of $\rho_\mu$ clearly shows that $\nu_\mu$ are
 disappearing on traversing the Earth, and hence must have oscillated
 into a combination of $\nu_e, \nu_\tau$ and $\nu_s$ (a singlet
 neutrino).\footnote{In this letter, we suppress our theoretical
 prejudice against singlet neutrinos, which apparently require either a
 new low energy scale of physics, or a non-minimal see-saw.} The present
 data strongly disfavors oscillations purely to $\nu_e$, but does not 
 distinguish
 between pure $\nu_\tau$ and $\nu_s$. A high statistics
 separation of the $\nu_\tau$ and $\nu_s$ modes requires event
 classes with substantial probabilities of being produced by
 neutral current interactions.

\end{itemize}

In this letter we study the neutrino oscillation signal in the up-down
ratio of inclusive multi-ring events, $\rho_{MR}$, both in the sub-
and multi-GeV data. These are all events in which neutrino
interactions produce two or more charged particles or photon showers
which lead to identified Cerenkov rings. 
The theoretical systematic uncertainties are 1--2\% from flux calculations, 
and 1--4\% from cross sections, and we expect the experimental
systematic uncertainties also to be small. The
large number of multi-ring events ensures small
statistical errors; for example, in the multi-GeV data
the number of multi-ring events is very similar to the sum of the
number of 1-ring $\mu$-like and PC events, so the statistical power of
$\rho_{MR}^{multi}$ will be the same as for $\rho_\mu$. For the sub-GeV data
the angular correlation will not be as strong as for the multi-GeV
case, but the multi-ring events have a higher neutral current
component, so that $\rho_{MR}^{sub}$ may have better capabilities to
separate $\nu_\tau$ from $\nu_s$. This separation can also be probed
via $\rho_{\pi^0}$, the up-down asymmetry of events with a single
$\pi^0$, as has been proposed \cite{DG}. While such events have a
higher neutral current component, a measurement of $\rho_{MR}^{sub}$ will
not require any particle identification, and will have better
statistics. 
It has been suggested that the $\nu_\tau$ and $\nu_s$ modes could be 
distinguished by studying the fraction of 2-ring events which result from 
neutral current $\pi^o$ production \cite{VS}. However, this method has large 
uncertainties due to the uncertainties in neutrino interaction cross sections.

\section{Up-down Ratios}

In this letter we present a simplified analysis for the up-down
ratios $\rho_i(\phi)$. We assume that, for some choice of the cone half
angle $\phi$, the downward going neutrinos are essentially
unoscillated, while the upward going neutrinos have completely
oscillated.\footnote{
Complete oscillation means that the relevant $\Delta m^2 L /E$ factors are 
sufficiently large that on averaging over $L/E$ the probabilities become  
independent of these factors. If this is not the case, $\rho_i$ is still given
by (\ref{eq:A}), with the probabilities now understood to be suitably averaged
over energy.} 
The measured zenith angle distributions make this
a reasonable assumption. We also assume that the flux ratio of
neutrinos produced in the atmosphere, $r = \nu_\mu / \nu_e$, is independent
of energy, at least for the energies which dominate a given 
class of events.\footnote{This is a better approximation for the sub-GeV
data than for the 
multi-GeV case. When $r$ has significant energy dependence, the up-down 
ratios are still given by (1), but with $r$ and $1/r$ suitably averaged.}
Then, for events of class $i$, which are induced by $\nu_e$ charged current,
$\nu_\mu$ charged current and neutral current interactions with
relative probabilities $f_e, f_\mu$ and $f_{NC}$
\begin{equation}
\rho_i = f_e (P_{ee} + r P_{\mu e}) + f_\mu \left( P_{\mu \mu} + { 1 \over r}
P_{e \mu}\right) + f_{NC} (1 - P_s)
\label{eq:A}
\end{equation}
where $P_{ij}$ is the probability for oscillation $\nu_i \rightarrow
\nu_j$ and $P_s = (P_{es} + r P_{\mu s}) / (1 + r)$. It is immediately
apparent that, within these approximations, up-down asymmetries can
only measure the combinations: $P_{ee} + r P_{\mu e},
P_{\mu \mu} + P_{e \mu}/r$ and $1 - P_s$; and the latter is only
probed if $f_{NC}$ is appreciable. The oscillation probabilities satisfy 
unitarity constraints, $\Sigma_j P_{ij} = 1$,
but the three combinations which can be measured 
remain independent. The fractions $f_e, f_\mu$ and
$f_{NC}$ can be obtained from the Monte Carlo results of the
Super-Kamiokande collaboration, and are shown in Table \ref{tb:f}.

\begin{table}[t]
 \renewcommand{\arraystretch}{1.5}
 \newcommand{\lw}[1]{\smash{\lower2.ex\hbox{#1}}}
 \begin{center}
  \begin{tabular}{llllrr} \hline\hline
    
                              & $f_e$ & $f_\mu$ & $f_{NC}$ & \#MC & \#data   \\
     \hline Fully Contained Sub-GeV (Analysis A) \\
     \hline 1 ring $e$-like   &  0.88 & 0.02    & 0.10   & 812.2 & 983       \\
            1 ring $\mu$-like &  0.005& 0.96    & 0.04   &1218.3 & 900      \\
             multi-ring        &  0.24 & 0.43    & 0.33  & 759.2 & 696      \\      
     \hline Fully Contained Multi-GeV \\
     \hline 1 ring $e$-like   &  0.84 & 0.07    & 0.09   & 182.7 & 218      \\
            1 ring $\mu$-like &  0.005& 0.99    & 0.005  & 229.0 & 176       \\
            multi-ring        &  0.30 & 0.55    & 0.15   & 433.7 & 398      \\
   \hline Partially Contained &  0.01 & 0.98    & 0.01   & 287.7 & 230      \\

\hline\hline
\end{tabular}
\end{center}
\caption{Data and Monte Carlo results reported for 25.5 kiloton-years, 
414 days of running time, by the Super-Kamiokande collaboration 
\cite{SKsub,SKmulti}, using the flux calculations of~Ref.\cite{Honda}.
For each class of event, the first three columns give the Monte Carlo 
results for the fractions of the events which are induced by $\nu_e$ charged 
current, $\nu_\mu$ charged current and neutral current interactions, the 
fourth column the expected number of events, and the fifth column the 
measured number of events.} 
\label{tb:f}
\end{table}

In the cases of oscillations between two flavors, such as 
$\nu_{\mu} \rightarrow \nu_{e}$, $\nu_{\mu} \rightarrow \nu_{\tau}$ 
or $\nu_{\mu} \rightarrow \nu_{s}$, with CP conservation, the 
expressions for the ratio simplify drastically and involve only one 
parameter, $P_{\mu\mu}$.  For $\nu_{\mu} \rightarrow \nu_{\tau}$ 
oscillations, $P_{ee} = 1$, $P_{\mu e} = P_{e \mu} = P_{s} = 0$, and 
$P_{\mu\mu} \leq 1$, giving the up-down ratio
\begin{equation}
        \rho_{i}^{(\tau)} = f_e  + f_\mu P_{\mu \mu} + f_{NC} .
        \label{eq:rho_tau}
\end{equation}
For $\nu_{\mu} \rightarrow \nu_{s}$ oscillation, $P_{ee} = 
1$, $P_{\mu e} = P_{e \mu} = 0$, $P_{\mu s} = 1 - P_{\mu \mu}$, and 
$P_{s} = r P_{\mu s}/(1+r)$, and we find
\begin{equation}
        \rho_i^{(s)} = f_e  + f_\mu P_{\mu \mu} + f_{NC} 
\frac{1 + rP_{\mu\mu}}{1+r} .
        \label{eq:rho_s}
\end{equation}
Finally, the case of $\nu_{\mu} \rightarrow \nu_{e}$ is the most 
complicated one.  Under the assumption of the CP conservation, $P_{\mu 
e} = P_{e\mu} = 1 - P_{\mu\mu}$, and $P_{ee} = P_{\mu\mu}$.  We have 
$P_{s} = 0$ in this case, and find
\begin{equation}
        \rho_i^{(e)} = f_e (r - (r-1) P_{\mu\mu}) + 
f_\mu \left({ 1 \over r} + { (r-1) 
        \over r} P_{\mu \mu}\right) + f_{NC}.
        \label{eq:rho_e}
\end{equation}
In these cases, measurements of the up-down ratios for various event classes, 
$i$, should yield the same value for $P_{\mu \mu}$. Values for $P_{\mu
\mu}$ extracted from $\rho_e^{sub,multi}$ and $\rho_\mu^{sub,multi}$
are shown in Table 2, using $r=2.15$ for sub-GeV data, and $r=3$ for
multi-GeV data~\cite{Honda}. 
For oscillations to $\nu_\tau$ or $\nu_s$, $P_{\mu
\mu}$ is essentially independent of $r$. For oscillations to $\nu_e$,
as $r$ is increased from 2.5 to 3.5, the value of $P_{\mu \mu}$ varies
from 0.20 to 0.32 (from $\rho_\mu^{multi}$), and from 1.13 to 1.08
(from $\rho_e^{multi}$). There is good consistency for oscillations of
$\nu_\mu$ to $\nu_\tau$ or $\nu_s$, but oscillations to $\nu_e$ are
excluded by the multi-GeV data at greater than $6 \sigma$, for any reasonable 
value for $r$.

For oscillations of $\nu_\mu$ to $\nu_\tau$ or $\nu_s$, the sub-GeV
data appears to give a somewhat larger value for $P_{\mu \mu}$ than
does the multi-GeV data. However, the measured value of
$\rho^{sub}(53^o)$ is larger than the true value, $\rho_T$, due to the
smearing effect from the poor angular correlation of the sub-GeV
data. This can be parameterized by a phenomenological parameter $D$:
$\rho^{sub}(53^o) = \rho_T + D (1- \rho_T)$. The central values of
$P_{\mu \mu}$ extracted from sub- and multi-GeV data coincide if $D \approx 
1/3$. This is consistent with the quoted angular resolution for the
sub-GeV data.

\begin{table}[t]
 \renewcommand{\arraystretch}{1.5}
 \newcommand{\lw}[1]{\smash{\lower2.ex\hbox{#1}}}
 \begin{center}
  \begin{tabular}{llll} \hline\hline
    
                              & $\nu_\mu \rightarrow \nu_\tau$ &
 $\nu_\mu \rightarrow \nu_s$ & $\nu_\mu \rightarrow \nu_e$   \\
\hline 
$\rho_\mu^{sub}(53^o)$&$0.67\pm 0.08^*$ & $0.68\pm 0.08^*$ & $0.38\pm 0.15^*$\\
$\rho_e^{sub}(53^o)$ & Consistent at $2 \sigma^*$ & Consistent at $2 \sigma^*$ 
& $0.77\pm 0.12^*$ \\
\hline
$\rho_\mu^{multi}(78^o)$& $0.52 \pm 0.07$ & $0.52 \pm 0.07$ & $0.24\pm 0.11$ \\
$\rho_e^{multi}(78^o)$ & Consistent at $1 \sigma$ &Consistent at $1 \sigma$
 & $1.10 \pm 0.08$ \\

\hline\hline
\end{tabular}
\end{center}
\caption{Values of $P_{\mu \mu}$ extracted from data on up-down ratios, 
assuming perfect angular correlation. $\rho_\mu$ includes both 1-ring 
$\mu$-like and PC events. $^*$ Only statistical uncertainties are included 
in the sub-GeV case.}
\label{tb:P}
\end{table}

To measure $\rho_{MR}$, it is necessary to define a direction for the
multi-ring events, and this can be done in many ways. In contrast to 
single-ring and PC events, it cannot be defined using one
particular ring; the direction must be defined 
in an inclusive way. The vertex should first be determined using the
standard method, then vectors from the vertex to individual hits in
the photomultiplier tubes (PMTs) can be drawn.  One way to define the
direction of the event is the sum of all the vectors weighted by the
corrected number of photo-electrons in the PMTs \cite{thesis}.  
An equivalent way is to use the
analog of the thrust variable in the QCD jet studies at $e^+ e^-$
colliders, 
\begin{equation}
  T = \mbox{max}_{\vec{n}} 
  \frac{ \sum_i |\vec{p}_i \cdot \vec{n}| }{\sum_i |\vec{p}_i|},
\end{equation}
where the maximum is obtained by varying a vector with a unit length 
$\vec{n}$.  The $\vec{n}$ which maximizes the thrust is the direction 
of a jet.  At water Cerenkov detectors, the momentum vector, $\vec{p}_i$, 
can be taken as the vector to each PMT weighted by the corrected number of 
photo-electrons in the PMT. This value is approximate, as it ignores the 
dependence of photo-electron production on particle type \cite{thesis}.
The advantage of using the thrust 
variable is that one can study the thrust distribution to see if 
Monte Carlo calculations give a reasonable agreement with the data.  Whatever 
definition of the direction of the event is employed, a detector 
simulation is necessary to study the correlation between the direction 
of the event and the original neutrino direction.

\section{Multi-GeV Data}

We propose an analysis of the zenith angle dependence in 
inclusive, multi-GeV, FC multi-ring events.  According to the Monte
Carlo study by the Super-Kamiokande Collaboration \cite{SKmulti} summarized in
Table~\ref{tb:f}, the FC multi-ring events are dominated by CC
events.  Since the multi-ring event sample is richer in multi-pion
production and hence to higher energy neutrinos than the single ring
event sample, the angular resolution is expected to be good.  We
actually do not know the breakdown of the FC multi-GeV multi-ring
event sample into quasi-elastic (QE), single pion, and multi-pion
production; the corresponding breakdown in the sub-GeV sample for
Kamiokande experiment by Kajita \cite{Kajita} shows that the QE contribution
is small (less than 5.9\% with $\nu_e$ CC, $\nu_\mu$ CC and NC
combined), and the rest is roughly equally divided between single pion
and multi-pion production.  In the multi-GeV sample, the multi-pion
production is expected to be more important, and can roughly be
approximated by the Deep Inelastic Scattering (DIS) processes, even though
the energy of the neutrino is still relatively low.  For the DIS
processes, the correlation between the neutrino
direction and the momentum of the hadronic system and the charged
lepton can be studied, and is of course perfect if one can observe all of 
the momenta.
Even though one misses both particles below the Cerenkov threshold and
neutrons, the correlation should be better than for QE events, where it
is 15--20\% (RMS) \cite{SKmulti}.  Therefore, we assume
in this paper that the FC multi-GeV multi-ring events have reasonable
angular resolution, at least as good as the QE events.  
We advocate evaluating $\rho_{MR}^{multi}(\phi)$ with a large half cone 
angle of $\phi = 78^o$, so that only about 20\% of the events are not used.

Once a direction of the event is defined, as outlined in the previous 
section, we can study the
zenith-angle dependence of the FC multi-GeV multi-ring events.  Since
more than half of the events are expected to be $\nu_\mu$ CC events,
the same anomaly seen in $\mu$-like one-ring events must also appear in
this zenith angle dependence.  In a 25.5 kt-years data sample, 398 events 
were observed, which gives roughly 11\% statistical uncertainty
in the up-down ratio.

The up-down ratio is also useful to study the nature of the
oscillation, to distinguish $\nu_\mu \rightarrow \nu_e$ oscillation
from $\nu_\mu \rightarrow \nu_\tau$ oscillation.
We denote the survival probability of $\nu_{\mu}$ as $P_{\mu\mu}$
as in the previous section.  In the 
case of $\nu_{\mu} \rightarrow \nu_{\tau}$ oscillation, the rate for upward 
going $\nu_{\mu}$ CC events is suppressed by the factor $P_{\mu\mu}$, while 
$\nu_{e}$ CC and NC event rates are unchanged.  
The resulting up-down ratio can be calculated using
Eq.~(\ref{eq:rho_tau}) and Table~\ref{tb:f},
\begin{equation}
  \rho_{MR}^{(\tau)multi} = 0.30 + 0.55 P_{\mu\mu} + 0.15.
\end{equation}
For $\nu_{\mu} \rightarrow \nu_{s}$ oscillation, Eq.~(\ref{eq:rho_s}) gives
\begin{equation}
  \rho_{MR}^{(s)multi} = 0.30 + 0.55 P_{\mu\mu} 
  + 0.15 \frac{1 + r P_{\mu\mu}}{1+r} .
\end{equation}
Here, the $\nu_\mu$ to $\nu_e$ flux ratio is roughly $r \approx 3$,
suitably averaged over a range of energies and zenith angles,
which can be studied in detail with their Monte Carlo analysis.
Finally, the case of $\nu_{\mu} \rightarrow \nu_{e}$ oscillation is 
more complicated.  The up-down ratio is
\begin{equation}
  \rho_{MR}^{(e)multi} = 0.30 (r - (r-1) P_{\mu\mu}) 
  + 0.55 \left({ 1 \over r} + { (r-1) \over r} P_{\mu \mu}\right) + 0.15 
\end{equation}
from Eq.~(\ref{eq:rho_e}).

One important question is the theoretical uncertainty in the
composition of the FC multi-GeV multi-ring events.  The thesis by
Shunsuke Kasuga \cite{thesis} varied the CC and NC cross sections for
sub-GeV data sample very conservatively by $\pm$30\% and $\pm$50\%,
respectively.  We take the same variation and determine the range
allowed for the up-down ratio as a function of $P_{\mu\mu}$.

\begin{figure}[tbp]
  \begin{center}
    \leavevmode
    \psfig{file=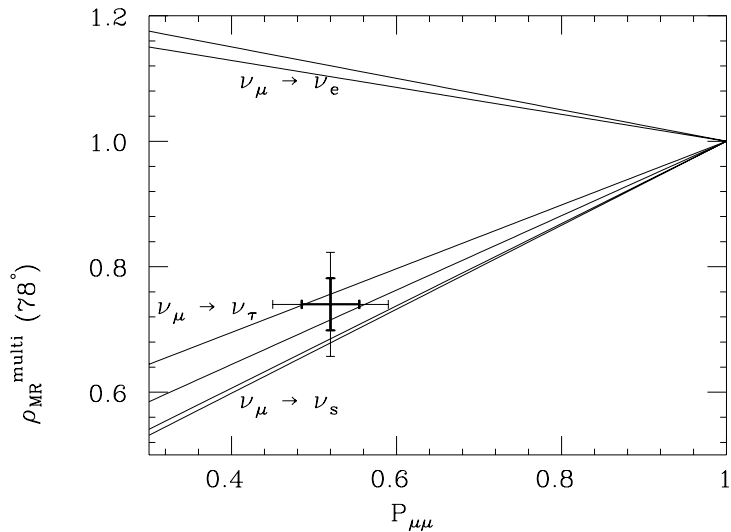,width=0.6\textwidth,angle=90}
    \caption{Up-down ratios in the FC multi-GeV multi-ring events for
      three oscillation scenarios.  The CC and NC cross sections are
      varied by $\pm$30\% and $\pm$50\%, respectively, to give bands for each
      cases.  The data point 
      with large error bars correctly reflects the observed up-down
      ratio in multi-GeV $\mu$-like events and its error in the
      horizontal direction, while the vertical position is yet to be
      measured.  The large vertical error bar is the expected statistical
      uncertainty using 414 days of data. The smaller error bars correspond 
      to four times more data -- about 5 years.}
    \label{fig:f}
  \end{center}
\end{figure}

Figure~\ref{fig:f} shows the up-down ratios predicted for the three
oscillation scenarios as a function of the $\nu_{\mu}$ survival
probability $P_{\mu\mu}$.  An important point is that the up-down
ratio is quite insensitive to the theoretical uncertainty in the cross
sections.  Another point is that three scenarios are relatively well
separated, even though $\nu_{\tau}$ and $\nu_{s}$ are 
somewhat close because of the small NC fraction.  The current up-down ratio
in the multi-GeV $\mu$-like events determine $P_{\mu\mu}$ with an
accuracy shown by the large error bar.  
The up-down ratio in the FC multi-GeV multi-ring events is
not reported; we simply took a number with an anticipated statistical
error.  It is clear that the current data sample can distinguish the
$\nu_{e}$ case and the other two.  The measurement of the up-down
ratio in this data sample will provide an independent
consistency check of the observed anomaly, without relying on any particle
identification.

%

\section{Sub-GeV Data}

The sub-GeV multi-ring data is potentially even more interesting than the 
multi-GeV data because it is rich in neutral current (NC) events, with
a fraction $f_{NC}$ of about a third.  
Therefore, this data sample has considerable power to 
discriminate between $\nu_{\mu} \rightarrow \nu_{s}$ oscillations and 
$\nu_{\mu} \rightarrow \nu_{\tau}$ oscillations.

The angular resolution is expected to be poorer than for the multi-GeV
data. However, judging from a Monte Carlo calculation of 
the zenith-angle distribution of the one-ring $\mu$-like events with
neutrino oscillations \cite{Kearns}, the zenith angular dependence is
not completely washed out; rather it is diluted only by $D \approx
0.4$.  The Super-Kamiokande
Collaboration quotes a mean angular correlation in one-ring events to
be 54\% for muons and 62\% for electrons \cite{SKsub}.  Therefore, it 
appears that the zenith-angle dependence or, in particular, the 
up-down ratio can be studied even with the sub-GeV data sample.  To 
be conservative, we study $\rho_{MR}^{sub}(\phi)$ with $\phi = 53^o$,
ie. we use only the up-most 
bin ($\cos \Theta < -0.6$) and the down-most bin ($\cos\Theta > 0.6$)
of the data. In this case, good angular resolution is not needed: 
$\Delta (\cos \Theta) \lesssim 2/5$, or $\Delta \Theta \lesssim 53^{\circ}$ is
sufficient, and is a reasonable expectation.

\begin{figure}[tbp]
  \begin{center}
    \leavevmode
    \psfig{file=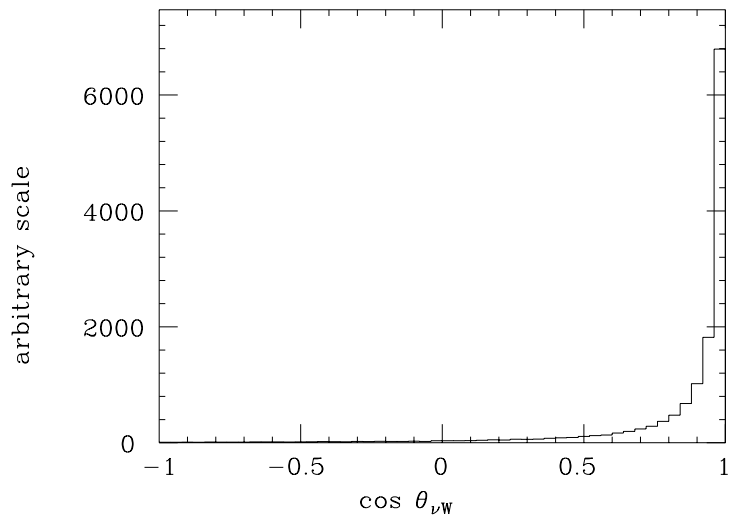,width=0.6\textwidth,angle=90}
        \caption{The angular correlation between the momentum of the
          neutrino and the hadronic system $W$ in DIS NC
          events.  The neutrino energy spectrum is taken from 
          the calculation of Honda
          {\it et al}\/ \cite{Honda}.}
        \label{fig:NCcorrelation}
      \end{center}
\end{figure}

The most interesting component of the sub-GeV, multi-ring 
events is the NC, which is made of single $\pi^{0}$ production and 
multi-pion production with almost an equal amount according to 
Kajita's table \cite{Kajita}.  The angular dependence in the single $\pi^{0}$ 
production was studied by Diwan and Goldhaber \cite{DG}, and is quite good. 
They proposed to study this mode exclusively by 
enhancing the angular resolution with a kinematical cut whose 
efficiency is lower than 30\%.  We, on the other hand, are interested 
in studying the multi-ring sample inclusively to collect maximum 
statistics.  Even without cuts, the contamination from the wrong bins 
is not large.  Furthermore, the multi-GeV
$\mu$-like data suggests  
that the zenith-angle distribution is more-or-less flat for the first 
two bins and the last two bins.  This also helps us to reduce the 
problem of poor angular resolution.  Finally, multi-pion 
production in the NC data is expected to show a good angular 
resolution, much better than the one-ring events.  The NC multi-pion
production in the sub-GeV  
category is coming from higher energy neutrinos than the single-ring 
events because (1) the loss of the lepton energy makes the NC events 
appear less energetic than they actually are, and (2) the cross 
section rises as a function of energy unlike the QE and single-pion 
processes.  We studied the angular correlation between the neutrino 
momentum and the momentum of the hadronic system numerically, and 
found indeed that the correlation is excellent 
(Fig.~\ref{fig:NCcorrelation}).  In reality, one cannot measure the 
total momentum of the hadronic system, because neutrons are not seen, and 
protons have a high energy threshold for Cerenkov radiation.  The true 
resolution needs to be studied with a full detector simulation.  
Still, it appears reasonable to assume that the angular resolution is 
sufficient for our analysis.

\begin{figure}[tbp]
  \begin{center}
    \leavevmode
    \psfig{file=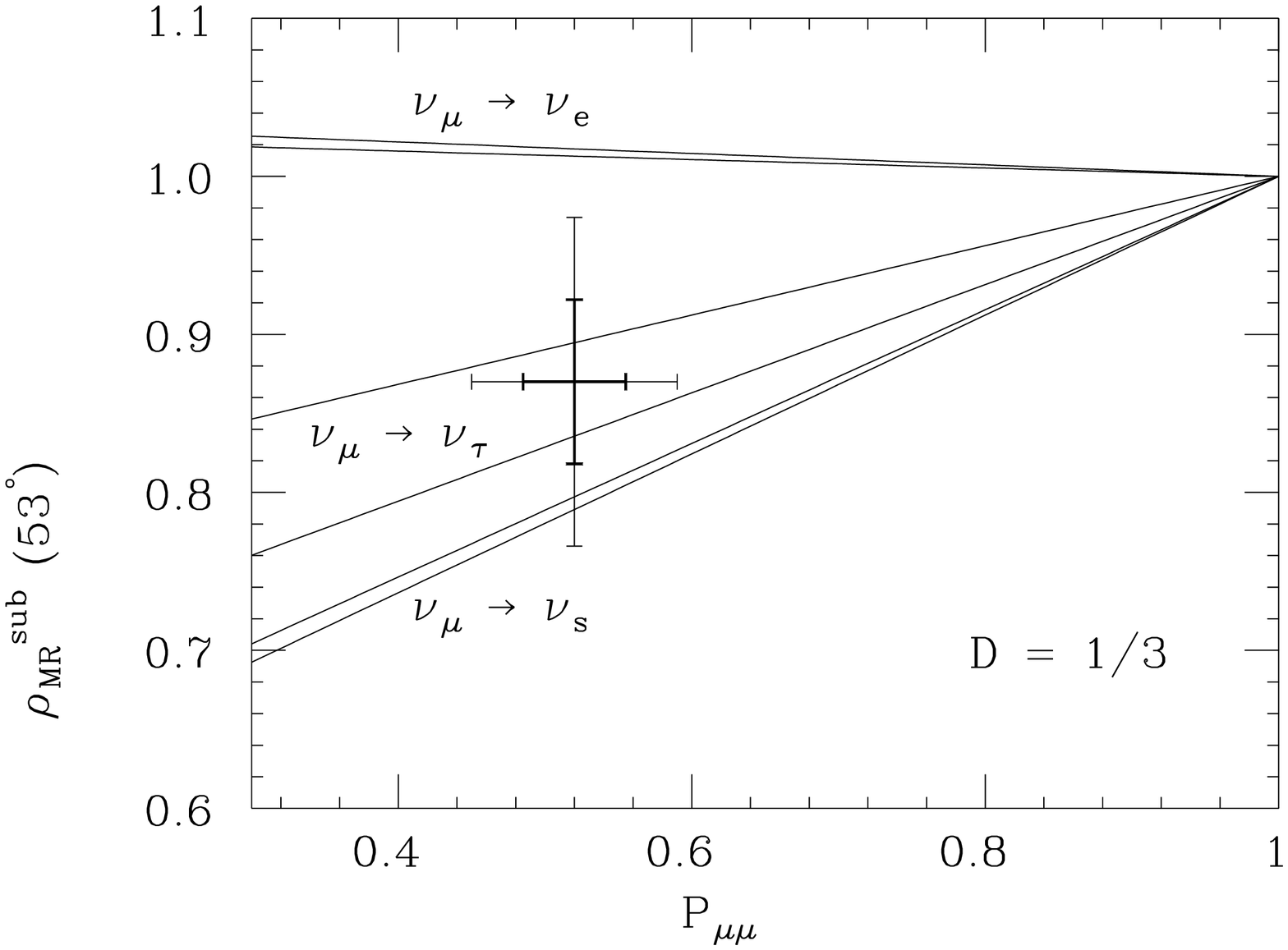,width=0.6\textwidth,angle=90}
        \caption{Up-down ratios in the FC sub-GeV multi-ring events for
          three oscillation scenarios.  The CC and NC cross sections
        are varied by $\pm$30\% and $\pm$50\%, respectively, to give bands for
          each cases.  The data point with large error bars 
          correctly reflects the observed
          up-down ratio in multi-GeV $\mu$-like events and its error
          in the horizontal direction, while the vertical position is
          yet to be measured.  The large vertical error bar is the expected
          statistical uncertainty using 414 days of data.  
          The smaller error bars correspond 
          to four times more data -- about 5 years. Only
          the upmost ($\cos \Theta < -0.6$) and downmost ($\cos\Theta
          > 0.6$) bins out of 5 are used, and a dilution of
          the ratio is included by taking $D=1/3$.}
        \label{fig:sub-updown}
        \end{center}
\end{figure}

Following the same analysis as for the multi-GeV data set, we show the 
expected behavior of the up-down ratio $\rho_{MR}^{sub}$ as a function of the 
$\nu_{\mu}$ survival probability $P_{\mu\mu}$ for three oscillation
scenarios.  By  
varying the CC and NC cross sections by $\pm$30\% and $\pm$50\%, respectively, 
the three scenarios give bands shown in the Figure 3.  Just like for the 
multi-GeV sample, the ratios are quite insensitive to a large 
variation of the cross sections.  The important point is the 
$\nu_{\tau}$ and $\nu_{s}$ scenarios are well separated.  This figure
is drawn with a dilution factor $D= 1/3$ in the up-down ratio, conservatively
allowing for a poor angular correlation.  In this plot, we assumed
that all of the events had the same dilution, which is probably a too
pessimistic assumption; the multi-ring NC events must have better
angular correlations than the QE events as discussed above.  

The published data set has 696 sub-GeV multi-ring events.  Because 
we conservatively use only the up-most and down-most bins, the 
expected current statistical error is about 12\%.  It is clear that 
oscillations to $\nu_{e}$ and to $\nu_{\tau}$ can be clearly separated.  In a 
few years, the $\nu_{\tau}$ and $\nu_{s}$ scenarios will be separated 
as well.

\section{Conclusions}

We have proposed a new test for atmospheric neutrino oscillations
using currently available data: the up-down ratio of FC, multi-ring,
inclusive events, $\rho_{MR}$. 
This method has good statistics and does not require particle identification.
We anticipate that, with 414 days of data, two
signals for neutrino oscillation could be obtained, one at the 
$3 \sigma$ level. This would provide an important, independent test of 
atmospheric neutrino oscillations. These measurements, one with sub-GeV data
and the other with multi-GeV data, could also confirm that $\nu_\mu$
oscillate to $\nu_\tau$ or $\nu_s$, and not to $\nu_e$. We believe
that $\rho_{MR}$, like $\rho_{e,\mu}$ are currently statistics
limited, so that the significance of these measurements will improve
with time. Combining $\rho_{MR}^{sub}$ and $\rho_{MR}^{multi}$ results
could allow a $2 \sigma$ separation of $\nu_\tau$ and $\nu_s$ modes
with 5 years of data. With 5 years data it is likely that $\Delta m^2$
can be reliably extracted from the zenith angle dependence of the
1-ring $e$-like, 1-ring $\mu$-like and PC events. It may then be
possible to increase the significance of the sub-GeV, multi-ring analysis
by including data from all zenith angles.
We believe that our sub-GeV analysis has been conservative: we have included 
a large dilution factor for angular correlations, we have taken very large 
uncertainties in the neutrino cross sections, and we have not attempted to 
optimize the cone angle $\phi$, so that it will be possible to improve the 
power of this technique.

Our analysis has largely been based on the assumption that the
oscillations involve only two flavors of neutrinos. Ultimately, it will
be important to perform a global analysis of the various up-down
ratios using the predictions of (1) for general neutrino
oscillations. The fit is overconstrained, as there are seven event
classes: $e$-like 1-ring, $\mu$-like 1-ring and multi-ring for the
sub-GeV data; FC $e$-like 1-ring, $\mu$-like 1-ring and multi-ring, as
well as PC events, for the multi-GeV data. The averaged neutrino flux
ratio, $r$, is close to 2 for the sub-GeV data, but 
for the multi-GeV data it is considerably
higher, and depends on the cone angle $\phi$. Hence, with several years
of data, it will be possible to fit the seven up-down ratios to obtain values 
for the five oscillation probabilities 
$P_{ee},P_{e\mu}, P_{\mu e}, P_{\mu\mu}$ and $P_s$.

\section*{Acknowledgements} 

This work was supported in part by the
U.S. Department of Energy under Contracts DE-AC03-76SF00098, in part
by the National Science Foundation under grant PHY-95-14797.  HM was
also supported by Alfred P. Sloan Foundation.

\end{document}